**Orientational ordering and layering of hard plates in narrow slit-like pores**

Sakine Mizani,[1] Roohollah Aliabadi,[2] Hamdollah Salehi,[1,*] and Szabolcs Varga[3]

[1]Department of Physics, Faculty of Science, Shahid Chamran University of Ahvaz, Ahvaz, Iran

[2]Department of Physics, Faculty of Science, Fasa University, 74617-81189 Fasa, Iran

[3]Institute of Physics and Mechatronics, University of Pannonia, P.O. Box 158, Veszprém, H-8201, Hungary

**Abstract**

We examine the ordering behavior of hard plate-like particle in a very narrow slit-like pore using the Parsons-Lee density functional theory and the restricted orientation approximation. We observe that the plates are orientationally ordered and align perpendicularly (face-on) to the walls at low densities, a first order layering transition occurs between uniaxial nematic structures having *n* and *n+1* layers at intermediate densities and even a phase transition between a monolayer with parallel (edge-on) orientational order and *n* layers with perpendicular one can be detected at high densities. In addition to this, the edge-on monolayer is usually biaxial nematic and a uniaxial-biaxial nematic phase transition can be also seen at very high densities.

**Introduction**

It is still challenging to understand the role of the shape, thickness-to-width, charge distribution and size-polydispersity of the colloidal particles on the stability of their different mesophases [1]. Over the years several hard body models including oblate and prolate have been devised and studied by simulation and theory [2]. Some important models of oblate (plate-like) particles are the disk [3-5], oblate spherocylinder [6,7], ring [8,9], cut sphere [10,11], sheet [12], lense [13], board [14,15] and rhombic platelet [16]. From the studies of these models, it is now well-understood that the key factor is the anisotropic shape in the formation of liquid crystalline states such as the nematic and columnar. The thin plate-like particles exhibit isotropic, nematic and columnar mesophases with increasing the density [2, 10], while the thicker ones may do form cubatic phase instead of nematic one [17,18]. If the plate-like particle is biaxial in shape such as the hard boards, a very complicated phase sequence emerges due to the stabilisation of the biaxial nematic phase and the appearance of the nematic-nematic phase transition [15]. Moreover, the surface charges on the plates can stabilize the hexatic [19] and smectic B [20,21] structures. The gravity and the size-polidispersity are the other important factors affecting the stability of the mesophases [22-26]. The phase behaviour of plate-like particles can be modified with addition of some depletion agents like the polymers [27] or mixing them with other colloidal particles having different size and shape such as spheres, rods and plates [28-30]. For example isotropic-isotropic and nematic-nematic demixing transitions can be observed in rod-plate [31,32], plate-plate [33] and plate-sphere mixtures [34,35].

---

[*] salehi_h@scu.ac.ir

The ordering of plate-like particles in the presence of a single wall and between two parallel walls have been studied with both theory and simulation [36-43]. The results of these studies can be summarized as follows: a) the plate particles wet the wall in face-on order (homeotropic anchoring), where the nematic director is perpendicular to the wall, b) the wall promotes the formation of nematic order, c) the capillary isotropic-nematic transition weakens with decreasing pore width and d) this first order transition terminates at a critical pore width which is in the order of $\sim 4D$ ($D$ is the diameter of the plate). However, the rods between two parallel hard walls exhibit planar anchoring, a wall induced uniaxial-biaxial surface ordering transition occurs and the capillary isotropic-nematic transition terminates at width of $\sim 2L$ ($L$ is the length of the rod)[44-49]. Furthermore, a layering transition between two smectic phases having *n* and *n*+1 layers exists in slit-like pores, where the anchoring of the rods is homeotropic [50-52]. Stiff ring polymers behave differently in the vicinity of hard walls, because they are adsorbed with edge-on order (planar anchoring) [53] and even a concentration induced planar-to-homeotropic anchoring transition can be detected [54]. This can be attributed to the penentrable nature of the rings, which can help to reduce the surface tension in homeotropic order. It is common in these confined studies that the pore is taken to be wide, which does not allow to study how the nature of orientational ordering changes between two- and three dimensions with widening the pore. Only few studies are devoted to examine the effect of out-of-plane fluctuations on the positional and orientational ordering transitions in very narrow pores [55-57]. In this regard, Khadilkar and Escobedo [58] studied the ordering of hard cubes in very narrow slit-like pore and observed layered structures and intermediate phases such as the buckled and rotator plastic phases. The effect of strong confinement was investigated for rod-like shapes as well, where a wall induced nematic ordering of non-mesogenic particles was detected [59]. To our best knowledge, the ordering properties of plate-like particles has not been studied in narrow slit-like pores.

In our present study, we investigate the effect of extreme confinement on the ordering properties of hard plates by placing the particles into very narrow slit-like pore. We focus on the effect of pore width and the plate's aspect ratio on the orientational order and surface adsorption. To avoid the relaxation to bulk properties in the middle of the pore and to avoid the formation of several mixed structures, the pore width is chosen such that the plates are allowed to form several layers in face-on (homeotropic) alignment, while only one layer can accommodate into the pore in edge-on (planar) one. We show that the strong confinement induces layered nematic structures and layering phase transitions between two uniaxial nematic phases having *n* and *n*+1 layers. Moreover, the surface anchoring changes from face-on to edge-on alignment and a biaxial nematic ordering becomes stable in edge-on order.

**Model and theory**

We place the rectangular plates with edge lengths $L$, $D$ and $D$ into a slit-like pore, where the confining walls are flat and parallel. We use the so-called three-state restricted orientation approximation, where the main symmetry axes of the particles can orient only along the $x$, $y$ and $z$ axes of the Cartesian coordinate system [60]. The schematic representation of the system, the molecular parameters and the possible orientations of the plates are shown in Figure 1. We examine the effect of varying aspect ratio ($L/D<1$) and the wall-to-wall distance ($H$) on the phase behavior of the plate-like particles. To do this we use the Parsons-Lee modification of the second virial density functional theory [61,62]. In this formalism the key quantity is the grand potential ($\Omega$), which is a functional of the local density ($\rho(1)$). In the three-state orientational model the system corresponds to a ternary mixture, where $x$, $y$ and $z$ orientations correspond to the components of the ternary mixture. Therefore the grand potential is a functional of the local densities of three orientations ($\rho_x(1)$, $\rho_y(1)$ and $\rho_z(1)$) as follows:

$$\beta\Omega[\rho] = \sum_{i=x,y,z} \int d(1)\rho_i(1)\left[\ln\rho_i(1) - 1 + \beta V_i^{ext}(1) - \beta\mu\right] - \frac{1}{2}c \sum_{i,j=x,y,z}\int d(1)\rho_i(1)\int d(2)\rho_j(2) \int f_{ij}^M(1,2),$$

(1)

where $\beta = 1/k_B T$ is the inverse temperature, $(1) = (x, y, z)$, $V_i^{ext}$ is the external potential between the plate with orientation $i$ and the walls, $\mu$ is the chemical potential, $c$ is the Parsons-Lee prefactor [63], $f_{ij}^M = \exp(-\beta u_{ij}(1,2)) - 1$ is the Mayer function and $u_{ij}(1,2)$ is the pair potential between particles with orientations $i$ and $j$ at the positions 1 and 2, respectively. As the interaction between the particles is hard repulsive, $f_{ij}^M$ is minus one for overlapping particles and zero otherwise. The external potential is infinite if the plate particle overlaps with the walls or the particle are outside the pore, while it is zero if the plate particle is inside the pore. This condition restricts the positions of the particles to be between the two hard walls. The functional minimization of Eq. (1) with respect to the component densities ($\rho_x(1), \rho_y(1)$ and $\rho_z(1)$) provides the equations for the equilibrium density profiles between the two parallel walls. As we do not intend to examine the crystalline structures in the *x-y* plane, the local densities of the three orientations depend on *z* coordinate only. The resulting set of equations for the local densities can be written as

$$\rho_k(z) = H\rho \frac{\exp\left[-c \sum_{i=x,y,z} \int dz' \rho_i(z') A_{ik}^{exc}(z,z')\right]}{\sum_{j=x,y,z} \int dz'' \exp\left[-c \sum_{i=x,y,z} \int dz' \rho_i(z') A_{ij}^{exc}(z',z'')\right]}, \qquad (2)$$

where $k=x$, $y$ and $z$, $\rho = N/V$ is the number density, $c=(1-3\eta/4)(1-\eta)^{-2}$, $\eta = \rho v_0$ is the packing fraction, $v_0 = D^2 L$ is the volume of the plate particle and $A_{ij}^{exc}$ is the excluded area between two hard plates with orientations *i* and *j*. The intervals of the integrations in *z* coordinate is restricted by the external potential. The details of the solution of the above set of equations is presented in our previous study [62]. After substitution of the solutions of Eq. (2) into Eq. (1) we get the equilibrium grand potential of the system. We also calculate the free energy of the system by the omission of the chemical potential term in Eq. (1). It is also possible to calculate the fraction of particles pointing into the direction *i* (*i=x, y* and *z*), which is defined by $X_i = N_i/N$, where $N_i$ is the number of plates in direction *i*. This can be obtained from the density profiles as follows

$$X_i = \frac{\int dz \rho_i(z)}{\sum_{j=x,y,z} \int dz \rho_j(z)}. \qquad (3)$$

In the case of first order phase transitions we determine the packing fractions of the coexisting phases $\alpha$ and $\beta$ from the equality of the pressures and chemical potentials, which are $P(\alpha) = P(\beta)$ and $\mu(\alpha) = \mu(\beta)$. In the next section we present our results in dimensionless units, where $D$ is taken to be the unit.

**Results and Discussion**

We study the orientational ordering and the layered structures of hard plate-like particles, which are confined into a pore by two parallel hard walls such a way that $L < H \leq L + D$. This condition allows the formation of a monolayer in edge-on orientation ($L$ side is parallel with the walls)) only, while several layers can accommodate into the pore in face-on direction ($L$ side is perpendicular to the walls). The upper limit of the number of layers is the integer of $H/L$ which cannot be more than 10 layers even for the lowest aspect ratio ($L/D$=0.1) we studied. If $L<H<D$ the plates are always in face-on direction to the walls and only layering phenomenon occurs between the two walls, because the particles cannot accommodate into the pore in edge-on direction. However, if $D<H<D+L$ both face-on and edge-on structures are feasible and the layered face-on structure can compete with an edge-on monolayer. This is due to the competition between different entropy contributions to maximize the available space for the particles in the pore. For example, the available volume of a plate is $A\,(H-D)$ and $A\,(H-L)$ for the edge-on monolayer and face-on layers, respectively, where $A$ is the surface area. Therefore, the available room (translational entropy) is always higher in face-on order than in edge-on one. Contrary to this, the particles exclude higher volumes from each other in face-on order than in edge-on one, i.e. the packing (excluded volume) entropy term supports the formation of edge-on monolayers. In addition to this even a competition between two layered structures having $n$ and $n+1$ layers may occur due to the interference between the wall induced oscillatory layered structures, which evolve from the opposite walls. In the case of $n$ layers, the translational entropy contribution is high, while the excluded volume one is low as the layers are wide. The opposite is true for $n+1$ layers, because the density peaks are sharper and the layers are thinner. In summary, we show together the competing face-on and edge-on structures in Figure 2.

We can gain some information about the high density structure of the system by examining the highest value of the packing fraction in different states. If the phase consists of $n$ layers in face-on order, the packing fraction ($\eta = \frac{N}{V} v_0$) can be factorised into two-dimensional (2D) and one-dimensional (1D) packing fractions as follows: $\eta = \eta_{2D} \eta_{1D}$, where $\eta_{2D} = \frac{ND^2}{nA}$ and $\eta_{1D} = \frac{nL}{H}$. Since the squares placed into square lattice can cover the 2D surface ($A$) perfectly, i.e. $\eta_{2D}(\max) = 1$, we get that $\eta(\max) = nL/H$ for $n$ layers. Since the maximum number of layers, which can fit into the pore, is the integer of $H/L$, we get that the close packing value of the packing fraction can be written as $\eta_{cp} = \text{int}\left(\frac{H}{L}\right)\frac{L}{H}$ in face-on order. In the case of edge-on order, we have only one layer, where $\eta_{2D} = \frac{NDL}{A}$ and $\eta_{1D} = \frac{D}{H}$. As the coverage of the 2D surface can be done perfectly without gaps, i.e. $\eta_{2D}(\max) = 1$, we get that $\eta_{cp} = \frac{D}{H}$. The comparison of the above close packing values of the two structures inform us that the stable phase is edge-on (face-on) in very dense phases, if $D/H$ is higher (lower) than $\text{int}\left(\frac{H}{L}\right)\frac{L}{H}$. We show later that the results of Eq. (2) for the stability of face-on and edge-on structures at high densities are consistent with the ordering direction of the close packing structure.

First we show the density profiles of the plates at three different densities in Figure 3, where $L/D$=0.6 and $H/D$=1.1. It can be seen that the local densities do not depend on $z$, because only one fluid layer is allowed to form in both face-on and edge-on orientations. At low densities the favoured structure is face-on order (Figure 3 (a)), because the available distance along $z$ direction is $0.5D$ in this state, while it is only $0.1D$ in edge-on one. At vanishing packing fraction (ideal gas limit: $\rho \to 0$) we get from Eqs. (2) and (3) that

$$\rho_x = \rho_y = \rho_z = \frac{H\rho}{2(H-D)+H-L}, \qquad X_x = X_y = \frac{H-D}{2(H-D)+H-L} \quad \text{and}$$

$$X_z = \frac{H-L}{2(H-D)+H-L}.$$ These equations show that the majority of the particles are aligned along the $z$ axis as $H$ goes to $D$ ($X_z \to 1$). In our special case these equations give that about 71% of the particles are in face-on order, while 29% of them are in edge-on one, which means that the phase is nematic even at vanishing density. Therefore the hard walls act like an external orientating field on the plate particles, where the nematic director is perpendicular to the walls (homeotropic ordering). However, the wall does not fix the direction of the nematic director, because more and more particles are ordered along the $x$ and $y$ directions with increasing density to minimize the excluded area between the plates. As a result an edge-on nematic order ($X_z < 0.5$) is obtained, which is uniaxial at $\eta = 0.6$ (Figure 3 (b)) and biaxial at $\eta = 0.8$ (Figure 3 (c)). In the uniaxial phase we can see that $\rho_x = \rho_y \gg \rho_z$, while $\rho_x > \rho_y \gg \rho_z$ in the biaxial nematic one. These two phases can be also considered as a planar order since the nematic director is in the $x$-$y$ plane. From these results, three different one-layer (1L) structures can be identified: a) one-layer face-on order (1LFO) if $X_z \geq 0.5$ and $X_x = X_y$ b) one-layer edge-on order (1LEO) if $X_z < 0.5$ and $X_x = X_y$ and c) one-layer biaxial order (1LBO) if $X_x > X_y > X_z$. The results of Eqs. (2) and (3) for the mole fractions are shown together in Figure 4 (a) at

$L/D$=0.6 and $H/D$=1.1, where the borders of different structures have been separated by vertical dashed lines. We show the stability regions of 1LFO, 1LEO and 1LBO structures in Figure 4 b) in $\eta$-$H/D$ plane for $L/D$=0.6. It is obvious that 1LFO structure can be destabilised with respect to 1LEO with increasing $H/D$, because the particles have more free volume (lower excluded area) in 1LEO order. The density of the uniaxial-biaxial transition (1LEO-1LBO) is only weakly affected by varying the pore width as this transition is an in-plane isotropic-nematic transition, which depends only on the number of particles being in the edge-on order. Here we note that 1LFO-1LEO structural change is not a true phase transition, because the thermodynamic quantities and the mole fractions are changing continuously with the density. In contrast to this, 1LEO-1LBO structural change is a second order phase transition. The 1LFO-1LEO change occurs in the density range of normal fluids, while the 1LEO-1LBO transition is probably pre-empted by freezing as $\eta_{1LEO-1LBO} \approx 0.8$. Even though our 1LBO phase is fluid, our prediction is right in that sense that the plate particles must order in edge-on direction and biaxial order to reach the close packing structure with increasing density. It is interesting that $\eta_{1LFO-1LEO}$ and $\eta_{1LEO-1LBO}$ do not exceed the maximum value of the packing fraction of face-on and edge-on structures, which are $L/H$ and $D/H$, respectively.

The structure of the plates changes substantially for $L/D$<0.5 because more than one layer can form between the two walls in face-on order. As a result the density profiles are inhomogeneous, the particles can adsorb to the walls in face-on order, the layered structures compete with each other and even face-on-to-edge-on orientational phase transition can occur. We show that this happens with plates if $L/D$=0.3. The density profiles of different structures are shown in Figure 5. We have obtained a two-layer face-on (2L) structure with some particles in the middle of the pore and a three-layer face-on (3L) at $\eta$=0.42 and $H/D$=1.04 (Figure 5 a) and b)). The main difference between these two structures is that the layers are thicker in the 2L face-on structure than in the 3L one. Since there are two solutions of Eq. (2) at the same inputs, where the free energy of 3L structure is lower than that of 2L one, there must be a phase transition between 2L and 3L face-on structures. Similarly we have found two solutions of Eq. (2) at higher packing fractions: 3L face-on and one-layer (1L) edge-on solutions (see Figure 5 c) and d)) at $\eta=0.51$ and 1L edge-on and four-layer (4L) face-on ones (see Figure 5 e) and f)) at $\eta=0.62$, where $H/D=1.2$. The observed 3L and 4L face-on structures are uniaxial as $\rho_x = \rho_y$, while the 1L edge-on one is biaxial as $\rho_x > \rho_y \gg \rho_z$. Using the phase equilibrium conditions we have determined the transition densities of the coexisting phases, which is presented in Figure 6 for $L/D=0.3$. Note that phase transition cannot occur below $H/D=0.9$, because the formation of the 2L face-on structure can develop continuously from a 1L edge-on one as the particles can adsorb easily to the walls. Since only 2L and 3L ordering is allowed to form for $0.9<H/D<1$ and the maximal packing fraction can be achieved with 3L structure, we find that a first order phase transition occurs between 2L and 3L structures for $H/D<1.05$. However, the 3L order can develop continuously from the 2L one for $H/D>1.05$. This is due to the fact that it is easier to accommodate three layers into the pore if the pore is wide enough. It can be also seen that the 3L phase becomes stable before we reach the maximum packing fraction of 2L phase (see the dashed curve in Figure 6). According to the close packing argument the plates should be in edge-on order at very high densities if $1<H/D<1.2$. In this region we observe a first order transition between a uniaxial 3L phase and a biaxial one-layer (1LBO) one with increasing density, where the orientational ordering changes from face-on to the edge-on direction. The high density stable phase is a four-layer (4L) structure for $1.2<H/D<1.3$, because $\eta_{\max}(4L) > \eta_{\max}(1LBO)$. Therefore an additional phase transition between 1LBO and 4L occurs at high densities. It is interesting that a face-on-edge-on-face-on ordering sequence can be observed with increasing density for $H/D=1.2$, i.e. the nematic director changes direction two times. One can see that more and more layers can accommodate into the pore with decreasing thickness of the plate particles because the maximum of the number of layers is int(*L/H*). We show two examples in Figure 7, where five-layer (5L) face-on structures change to six-layer (6L) face-on one for *L/D*=0.1. One can see that the formation of an extra layer in the middle of the pore can be achieved easily at *H/D*=0.775 (Figure 7 a) and b)) as the middle thick layer in 5L structure can split into two layers. This is not the case in a narrower pore (Figure 7 c) and d)), where the accommodation with 6 layers is harder as the layers are already thin in the 5 layer structure. As the sharpening peaks increase substantially the loss in the translational entropy and the packing entropy contribution increases from the lowering excluded areas with the formation of an extra layer, we encounter the competition of these two entropies and a first order layering transition between *n* and *n*+1 layers takes place. We show the resulting layering phase diagrams for different aspect ratios in Figure 8. Since maximum five layers can accommodate into the widest pore (*H/D*=1.2) for *L/D*=0.2, the following layering transitions may emerge in face-on order: 2L-3L and 3L-4L and 4L-5L. This is shown in Figure 8 (a), where the curves of maximal packing fractions of 2L and 3L structures can also be seen. Note that the maximal packing fraction of *n* layers is always above the *n*-*n*+1 layering transition. We have not observed 1LBO order because the close packing structure is degenerate, i.e. both the face-on and the edge-on orders produce the same maximal packing fraction. In addition to this, it is favourable to stay in layered structure, because the particles can access larger part of the space. Similar phase diagrams are obtained for *L/D*=0.15 and *L/D*=0.1, where the maximum of the number of layers are seven and ten, respectively. We can see that the number of layers increases with widening the pore. It is interesting that only one layering transition occurs at a given *H/D* and all first order layering transitions terminate at almost the same value of packing fraction ($\eta \approx 0.41$). At lower packing fractions, the *n*-to-*n*+1 layering are continuous structural change and the layered structures can develop continuously from the adsorbed layers of the walls. Our results show that even if there are more phase transitions with decreasing *L/D*, the phase diagram is less complicate, because the 1LBO order is destabilized and only the competition between *n* and *n*+1 layered structures survives. In order to make the phase diagram more complex, the pore should be wider in such an amount that mixed structures with both edge-on and face-on orders can also form.

**Conclusions**

We have studied the effect of the pore width and aspect ratio on the ordering properties of hard plate-like particles, which are placed between two parallel hard walls. Using the Parsons-Lee density functional theory in restricted orientation approximation, we have observed that the wall-particle hard interaction induces a uniaxial nematic order with strong adsorption at the walls. The nematic director is perpendicular to the walls (face-on order), i.e. the ordering can be considered as homeotropic. This kind of ordering maximizes the available room for the plates between the walls and the strong surface adsorption reduces the excluded volume cost between the particles. The adsorbed layers at the surfaces are uniaxial, i.e. only tetratic and solid phases can emerge in the adsorption layer if the surface density exceeds the transition densities of two-dimensional hard squares [64,65]. These possible in-plane orderings are not taken into account in our formalism. The number of layers between two parallel walls is an integer of *H/L* in face-on (homeotropic) order, while only one layer is allowed to form in edge-on (planar) order as *L<H<L+D*. As the walls are very close to each other (quasi-two-dimensional system), the wall induced face-on nematic fluid is inhomogeneous even in the middle of the pore, because the system cannot relax to the bulk values. Therefore continuous and first order layering transitions can occur between layered nematic fluids, where the two fluids have *n* and *n*+1 layers. If the pore is enough wide, the formation of a new layer can be realized continuously with a split of a wide fluid layer into two ones or with a formation of a new peak in the middle of the pore. However, the formation of a new layer in face-on order is accompanied by high translational entropy cost in several cases, which results in a first order layering phase transition. The edge-on nematic structure corresponds practically to a two-dimensional (2D) fluid of hard rectangles, where there are only few plates in face-on order. This phase is homogeneous and a 2D isotropic-nemtic phase transition may occur with increasing density, which corresponds to a uniaxial nematic-biaxial nematic phase transition. The biaxial nematic ordering is observed only in fluids of weakly anisotropic particles, where the uniaxial-biaxial transition density is very high, while it is not observed in fluids of strongly anisotropic particles. This is due to the fact that the face-on order becomes more stable than the edge-on one at a given pore width, because the available distance between the two walls is increasing in face-on order ($H-L$), while it does not change in edge-on one ($H-D$) as $L$ goes to zero. As a result the biaxial ordering is suppressed with $L/D \to 0$ and several layering transitions can be detected in face-on order. The first layering transition occurs between fluids having two and three layers, while the last layering transition takes place between two fluids with $n-1$ and $n$ layers with increasing pore width, where $n=\text{int}(H_{max}/L)$ and $H_{max}=L+D$. In the special case of infinitely thin plates (platelet) the number of layering transitions diverges with the pore width as $n$ goes to infinity. The observed layering transitions are not affected by a first order capillary nematisation transition because a capillary critical point emerges at the pore width of $\sim 4\,D$, which is higher than $H_{max}$.

It is worth to compare the ordering properties of confined plate-like and rod-like particles as they behave differently. Even though the rods exhibit also strong adsorption at the walls, the rod's long axis like to be parallel with the walls and form a planar layer. If the surface density of the adsorbed particles at the walls exceed a certain density, a surface induced uniaxial nematic-biaxial nematic transition emerges in the vicinity of walls, which corresponds to isotropic-nematic transition of 2D hard rods [44]. The layering transition happens mostly between two biaxial nematic fluids with $n$ and $n+1$ layers where the transition densities are above the density of the orientational ordering transition [66]. These results differ sharply from the plates, where the ordering is homeotropic, the uniaxial layering transition happens at intermediate densities and the biaxial nematic ordering may occur at very high densities. In addition to this the period of the layered structure is in the order of $D$ for rods while it is proportional to $L$ for plates.

Our results should be considered with some reservation at high densities, because the in-plane positional and orientation orderings are not taken into account. In addition to this, our mean-field type theory has the property that it exaggerate slightly the order of the phase transitions and overestimate the stability of the ordered phases. Therefore new simulation studies and experiments would be useful to justify our findings. However, our theory predicts correctly that the layering must occur with increasing density and it also produces the right ordering direction at very high density. In accordance with our findings, a recent MC simulation study found that the system of confined hard squares exhibits a weak layering transition [67]. In order to improve the reliability of the present theory for confined plate-like particles, the dimensional cross-over between 2D and 3D systems should be incorporated correctly. Along this line the fundamental measure density functional theory can be a step ahead [68-73], which proved accurate for infinitely thin plates [74].

Finally, we mention that attractive interactions can also play crucial rule in the stability of the different structures. For example the adding special square-well pair potentials to the hard-body interactions, even the vapour-liquid transition of the laponite plate-like particles can be described correctly [75,76].

ACKNOWLEDGMENTS

H.S., S.M. and R.A. thank Shahid Chamran University of Ahvaz and Fasa University for supporting the research and providing computing facilities. S.M. also appreciates Institute of Physics and Mechatronics, University of Pannonia, Hungary. S.V. acknowledges the financial support of the National Research, Development, and Innovation Office (Grant No. NKFIH K124353).

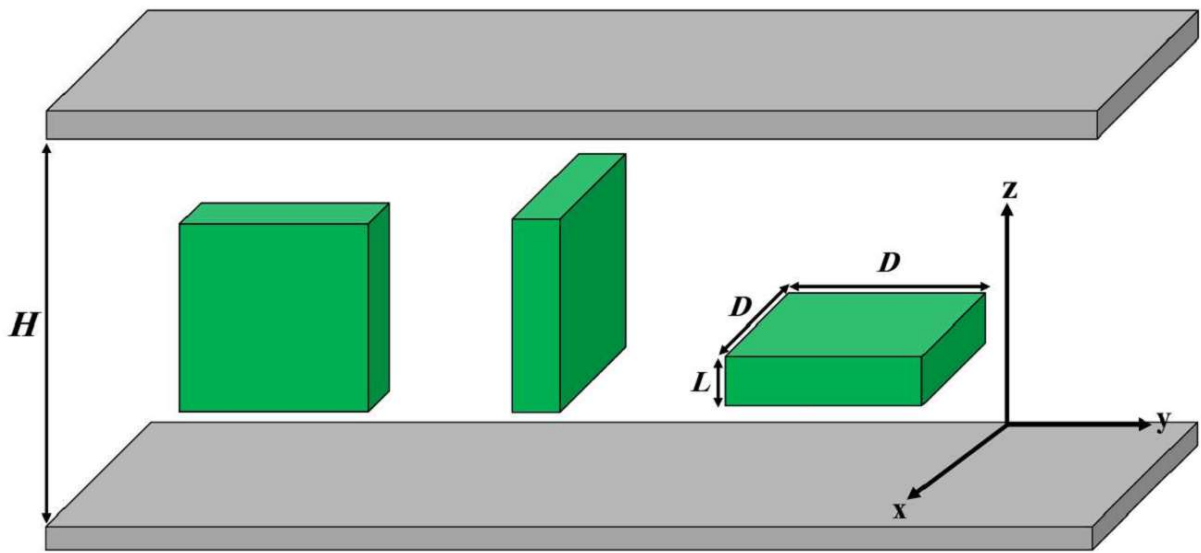

**Figure 1:** The three possible orientations of the hard plates between two parallel hard walls. The plate is oriented along the *x*, *y* and *z* axes from left to right. The *x* and *y* orientations are edge-on, while the *z* orientation is face-on to the walls. *H* is the pore width, *L* and *D* are the side lengths of the plate.

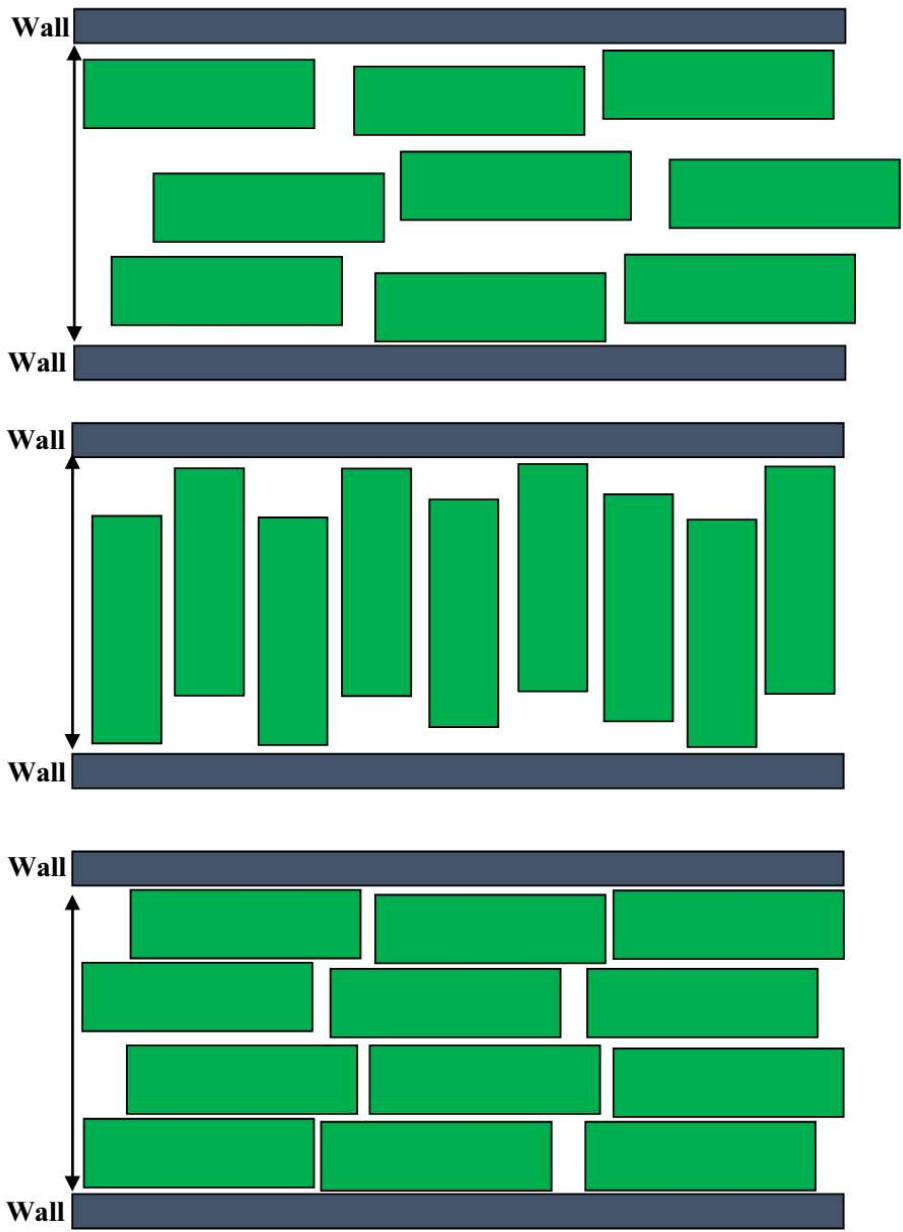

**Figure 2:** Two-dimensional representation of the possible structures of hard plates: a face-on ordering of the plates with the walls in three layers (upper panel), an edge-on ordering of the plates with the walls in one layer (middle panel) and a face-on ordering of the plates with the walls in four layers (lower panel). Here we use the following values: *L/D*=0.3 and *H/D*=1.3

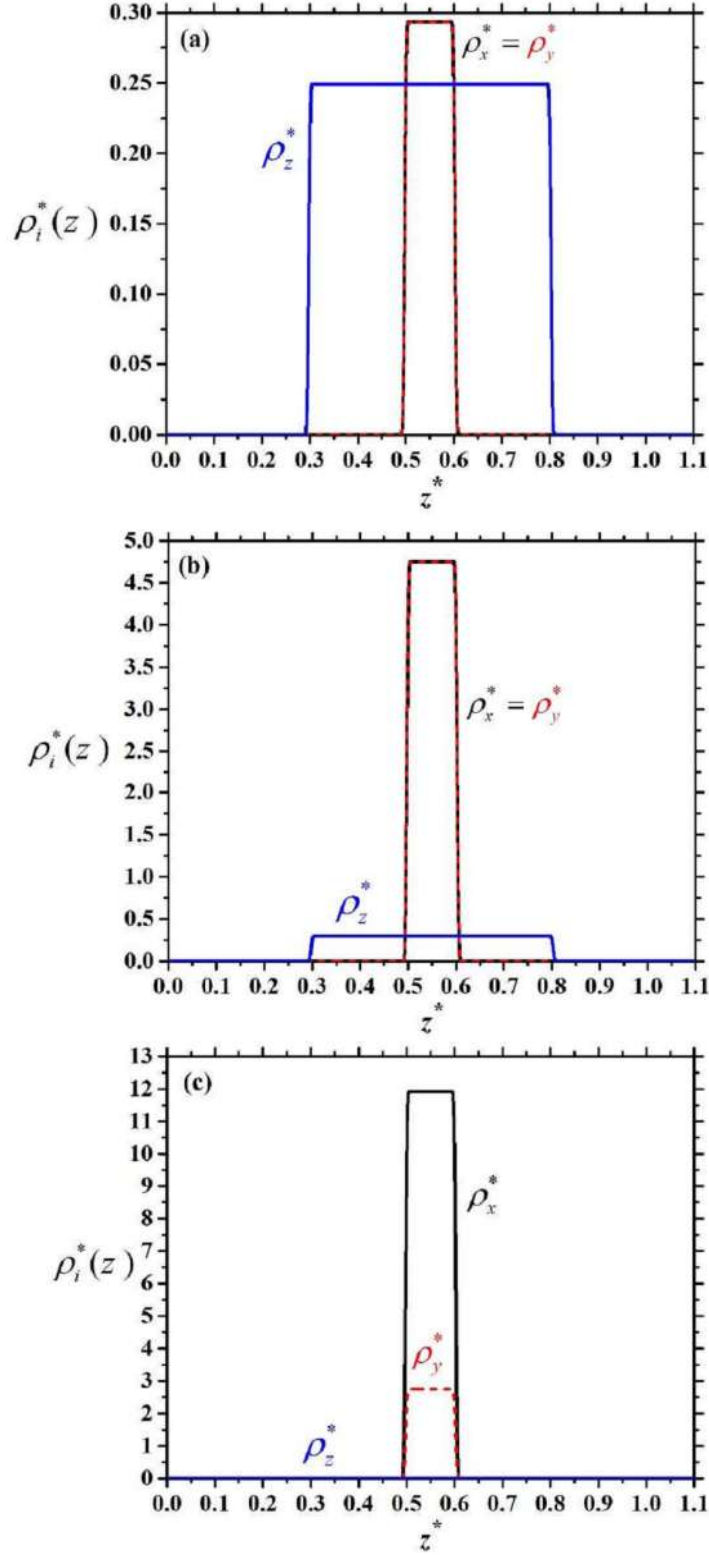

**Figure 3:** Density profiles of the plates for the three possible orientations for *L/D*=0.6 and *H/D*=1.1: (a) face-on order at $\eta=0.1$ as the majority of the particles are face-on to the walls, (b) uniaxial edge-on order at $\eta=0.6$ as $\rho_x=\rho_y$ and $X_z<1/2$, and (c) biaxial edge-on order

at $\eta = 0.8$ as $\rho_x \neq \rho_y$ and $X_z < 1/2$. The quantities are dimensionless, i.e. $z^* = z/D$ $\rho_i^* = \rho_i D^3$.

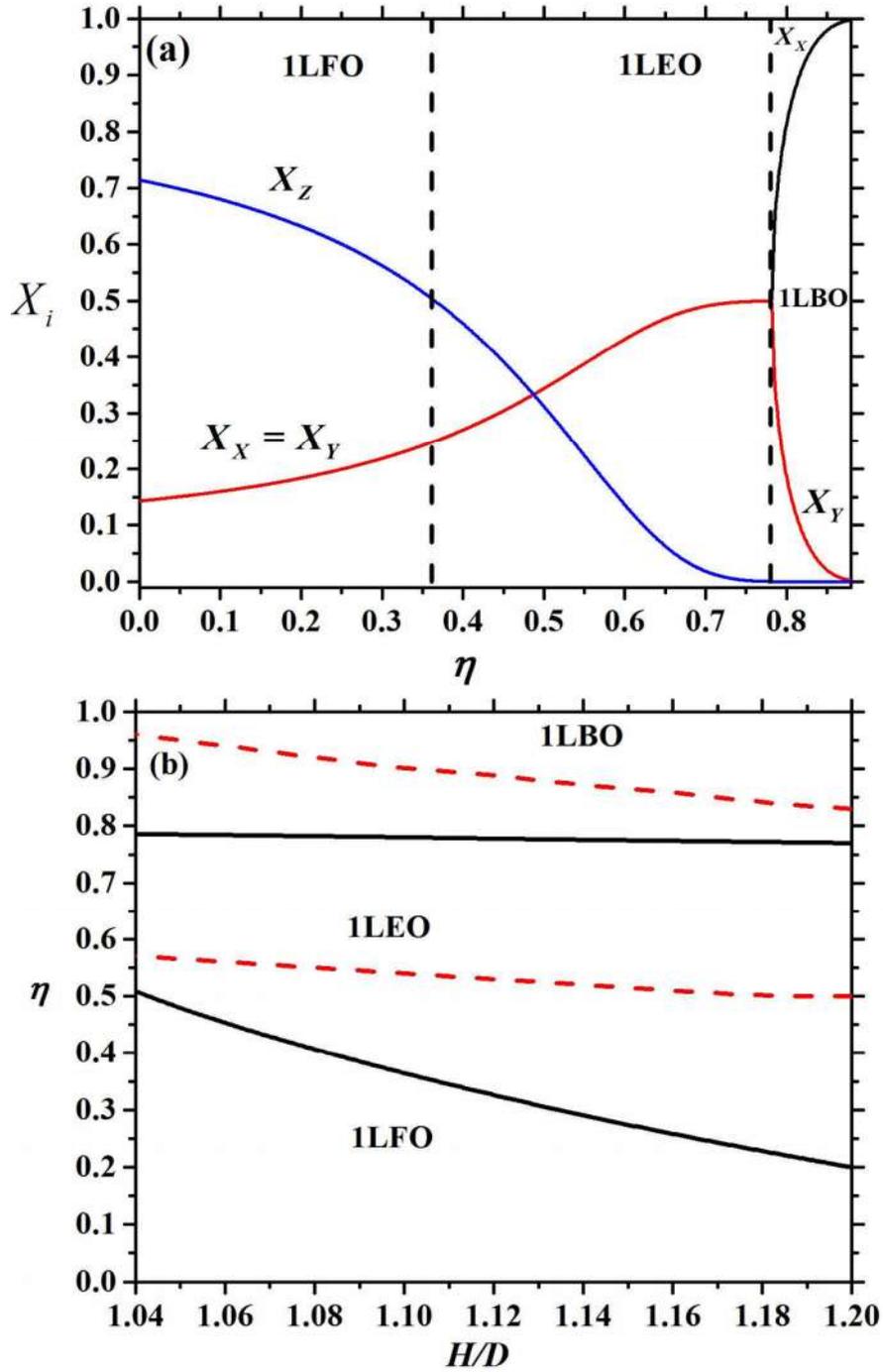

**Figure 4:** Phase diagram of confined hard plates with *L/D*=0.6. (a) Mole fractions of the plates in the three possible orientations as a function of packing fraction. (b) Borders of the observed

phases in packing fraction-pore width plane. The following phases are observed: one-layer face-on (1LFO), one-layer edge-on uniaxial (1LEO) and one-layer edge-on biaxial order (1LBO). The vertical dashed lines delimit the structures from each other in (a), while the lower and upper dashed curves represent the maximal packing fraction of face-on and edge-on monolayers in (b), respectively.

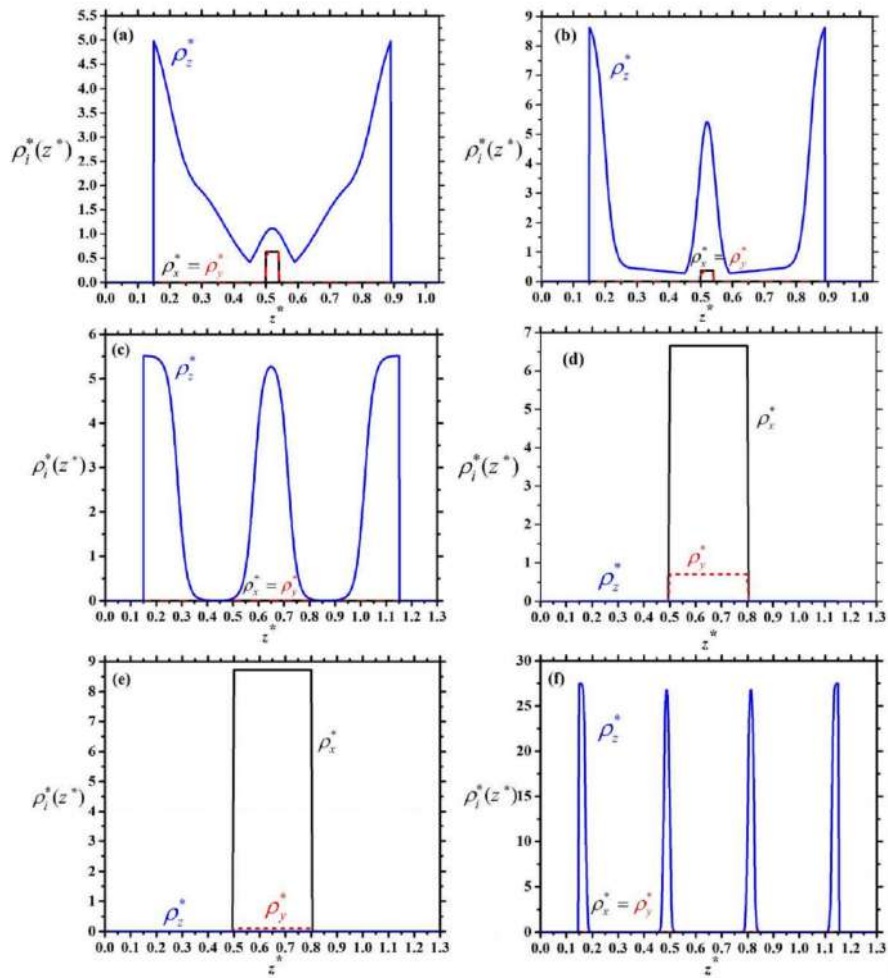

**Figure 5**: Density profiles of hard plates with $L/D=0.3$: (a) and b) two- and three-layer structures in face-on order at $\eta=0.42$ and $H/D=1.04$, (c) and (d) three-layer structure in face-on and one-layer biaxial in edge-on order at $\eta=0.51$ and $H/D=1.3$ and (e) and (f) one-layer edge-on biaxial and four-layer structure in face-on order at $\eta=0.62$ and $H/D=1.3$. The quantities are dimensionless, i.e. $z^* = z/D$ $\rho_i^* = \rho_i D^3$.

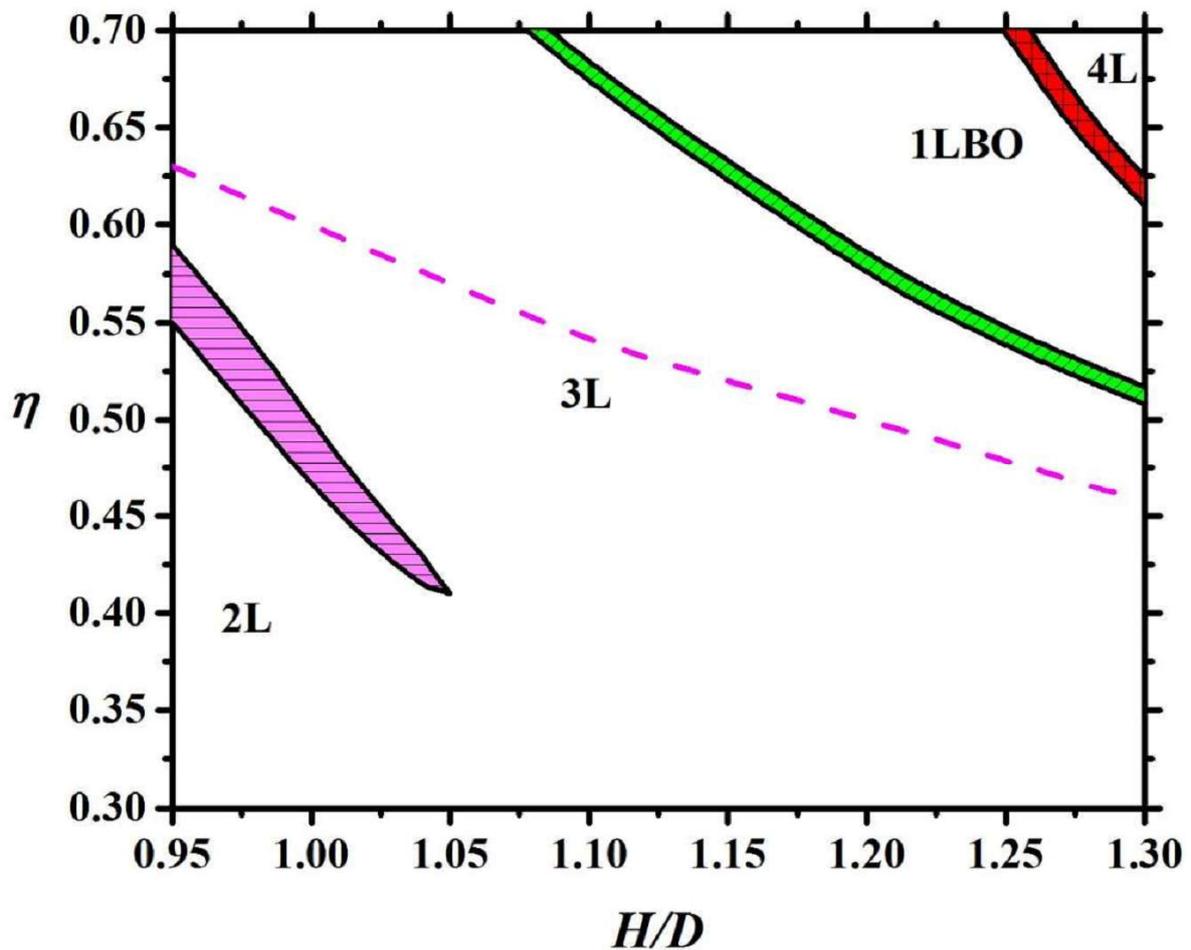

**Figure 6:** Phase diagram of confined hard plates with *L/D*=0.3 in packing fraction-pore width plane. The following structures are observed: face-on bilayer (2L), three-layer (3L) face-on, four-layer (4L) face-on and biaxial edge-on monolayer (1LBO). The biphasic regions are shaded. The dashed curve shows the maximal packing fraction of 2L structure.

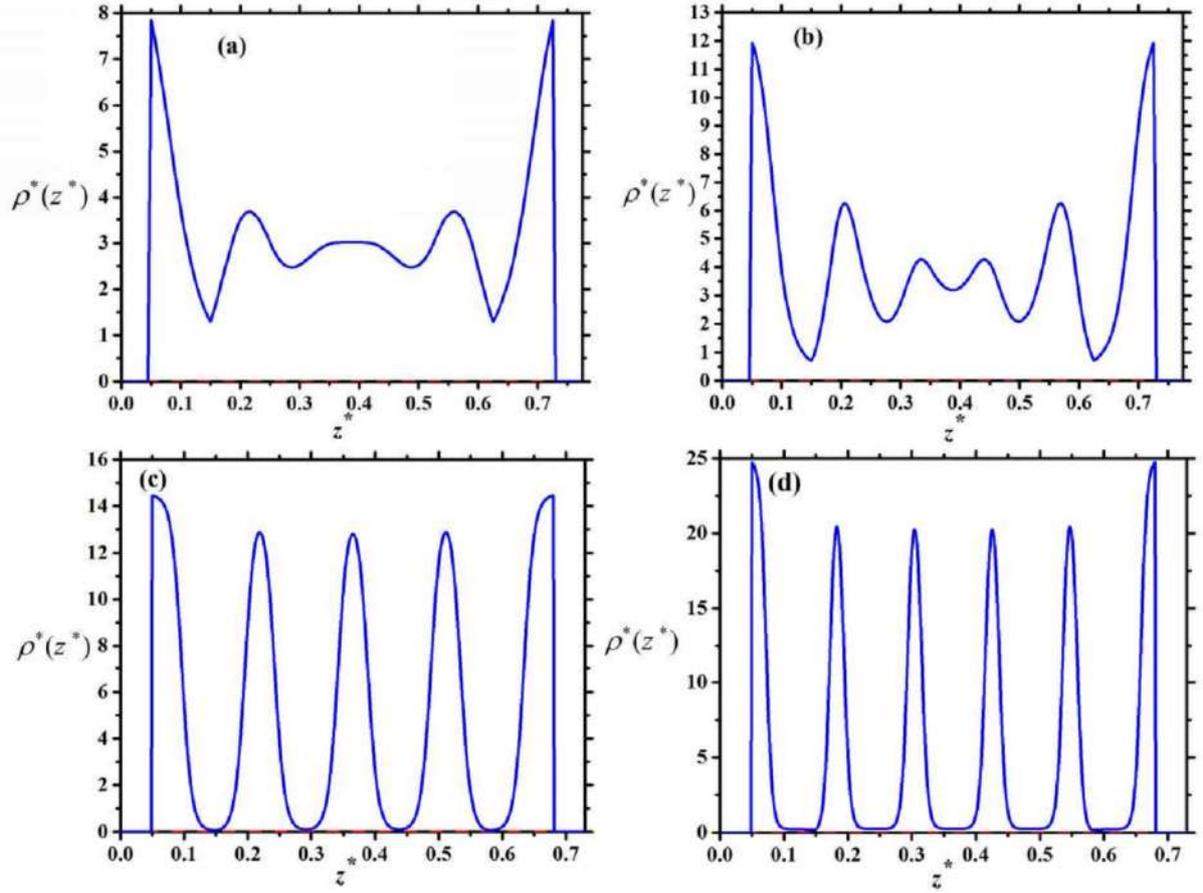

**Figure 7**: Density profiles of hard plates with $L/D$=0.1: (a) five-layer face-on structure at $H/D$=0.775 and $\eta$=0.3, (b) six-layer face-on structure at $H/D$=0.775 and $\eta$=0.34, (c) five-layer face-on structure at $H/D$=0.73 and $\eta$=0.455 and (d) six-layer face-on structure at $H/D$=0.73 and $\eta$=0.455. The quantities are dimensionless, i.e. $z^* = z/D$ and $\rho_z^* = \rho_z D^3$. At these molecular parameters $\rho_x^*(z^*) = \rho_y^*(z^*) = 0$ and $\rho_z^*(z) = \rho^*(z^*)$.

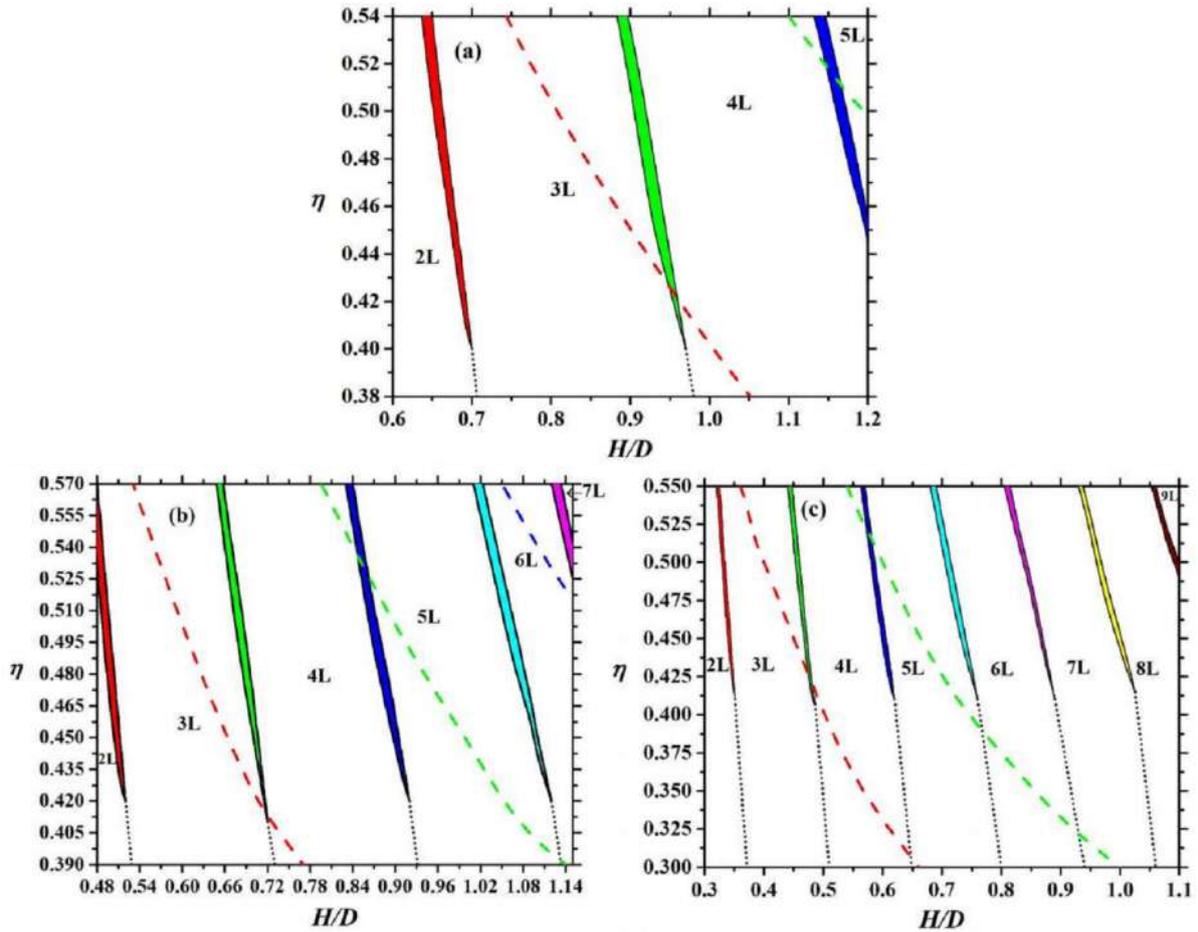

**Figure 8:** Phase diagram of confined hard plates with *L/D*=0.2 (a), 0.15 (b) and 0.1 (c) in packing fraction-pore width plane. Layered structures in face-on order with the walls are observed, where the number of layers is between 2 and 9 (2L,…,9L). The layering transition occurs between structures having *n* and *n*+1 layers. The biphasic regions of the layering transitions are shaded. The dashed curves show the maximal packing fraction of 2L, 3L and 4L structures from left to right. The dotted curve represents the continuous structural change from *n* to *n*+1 layers, where *n*=2,3,…7.